\documentclass[onecolumn,amsmath,amssymb,10pt,aps]{revtex4}
  
\usepackage[utf8]{inputenc}
\usepackage[T1]{fontenc}

\usepackage{amsmath}
\usepackage{amsfonts}
\usepackage{graphicx}
\usepackage{yfonts}
\usepackage{color}
\usepackage[normalem]{ulem}
\usepackage{amsthm}
\usepackage{bm}
\usepackage{bbm}
\usepackage{mathtools}
\usepackage{array}
\usepackage{placeins}
\usepackage{enumerate}
\usepackage{tikz}

\newcommand{\der}{\,\mathrm{d}}
\newcommand\bigforall{\mbox{\Large $\mathsurround=1pt\forall$}}

\def\<{\langle}
\def\>{\rangle}
\newcommand{\tr}{\mathrm{Tr}}
\newcommand{\Tr}{\mathrm{Tr}}
\def\oper{{\mathchoice{\rm 1\mskip-4mu l}{\rm 1\mskip-4mu l}
{\rm 1\mskip-4.5mu l}{\rm 1\mskip-5mu l}}}
\DeclareMathAlphabet\mathbfcal{OMS}{cmsy}{b}{n}

\mathchardef\mhyphen="2D 

\begin{document}

\title{Geometry of phase-covariant qubit channels}

\author{Katarzyna Siudzi\'{n}ska}
\affiliation{Institute of Physics, Faculty of Physics, Astronomy and Informatics \\  Nicolaus Copernicus University in Toru\'{n}, ul. Grudzi\k{a}dzka 5, 87--100 Toru\'{n}, Poland}

\begin{abstract}
We analyze the geometry on the space of non-unital phase-covariant qubit maps. Using the corresponding Choi-Jamio{\l}kowski states, we derive the Hilbert-Schmidt line and volume elements using the channel eigenvalues together with the parameter that characterizes non-unitality. We find the shapes and analytically compute the volumes of phase-covariant channels, in particular  entanglement breaking and obtainable with time-local generators.
\end{abstract}

\flushbottom

\maketitle

\thispagestyle{empty}

\section{Introduction}

In the recent years, non-unital channels have been receiving increasing attention. Of particular interest are phase-covariant qubit channels, where the non-unitality property is controlled with a single parameter. First master equations for phase-covariant dynamical maps were phenomenological in nature, and they were used to describe thermalization and dephasing processes beyond the Markovian approximation \cite{PC1}. Later, a microscopic derivation was presented, modelled with weakly-coupled spin-bosons under the secular approximation \cite{PC3}. During futher research, a connection was found between the population monotonicity, coherence monotonicity, and Markovianity of evolution \cite{PC2}. Due to high symmetries of phase-covariant channels, it was possible to estimate the upper and lower bounds of their classical capacity \cite{capacity_non-unital,experimental_capacity}. In the theory of open quantum systems, the evolution they provide was analyzed in terms of non-Markovianity \cite{phase-cov,ENM_non-unital,CC_GAD} and quantum speed limit \cite{QSL}.

However, many questions that have been answered for unital qubit maps are still open for non-unital channels. Among such open problems, there is their geometrical characterization, like the shape and size of the space of non-unital maps, or the analysis of the underlying geometrical structures. For the Pauli channels, which are unital qubit maps, Jagadish et al. proposed a measure, under which the volumes of positive and completely positive, trace-preserving maps were analyzed \cite{Jagadish}. However, it was also stated that the same measure cannot be used for non-unital channels \cite{Jagadish5}. More results for unital channels were obtained using the Lebesgue measure \cite{Lovas} and the Hilbert-Schmidt metric \cite{Pauli_volume,GGPC_volume,PV}. The corresponding dynamical maps have been also studied. The geometry of Markovian semigroups was considered \cite{Filippov,Karol_Pauli,Karol_Weyl}, as well as the volumes of non-Markovian dynamical maps under mixing semigroups \cite{Jagadish2} and their generalizations \cite{Jagadish3}. Recently, fractions of invertible maps obtained through mixing non-invertible maps have also been analyzed \cite{invertibility_measure}.

This paper is a continuation of refs. \cite{Pauli_volume,PV}, where we analyzed the geometry of Pauli channels. Using the Choi-Jamio{\l}kowski isomorphism, we introduced the Hilbert-Schmidt metric on the space of trace-preserving quantum maps. Through integration of the associated volume elements, we were able to calculate the relative volumes of completely positive maps, entanglement breaking channels, divisible maps, and quantum maps obtainable with time-local generators. This was accomplished for general Pauli channels, characterized by three distinct eigenvalues, as well as channels that are non-invertible or have degenerate eigenvalues. In this work, we generalize our results to non-unital channels. We derive the geometrical structures on the manifold of phase-covariant maps and compute the volumes of positive, trace-preserving maps. Next, we consider the relative volumes between selected classes of important quantum channels. The shapes and sizes of the channel spaces are graphically presented.

\section{Phase-covariant qubit channels}

Consider a qubit evolution that combines pure dephasing, energy emission, and energy absorption. It can be described by phase-covariant channels, which are completely positive, trace-preserving maps $\Lambda$ covariant with respect to a unitary transformation $U(\phi)=\exp(-i\sigma_3\phi)$, $\phi\in\mathbb{R}$. In other words,
\begin{equation}\label{cov_def}
\bigforall_{X\in\mathcal{B}(\mathcal{H}),\,\phi\in\mathbb{R}}\quad
\Lambda\big[U(\phi)X U^\dagger(\phi)\big] = U(\phi)\Lambda[X]U^\dagger(\phi).
\end{equation}
The most general form of a quantum map that satisfies eq. (\ref{cov_def}) reads \cite{phase-cov-PRL,phase-cov}
\begin{equation}
\Lambda[X]=\frac 12 \left[(\mathbb{I}+\lambda_\ast\sigma_3)\tr X+\lambda_1\sigma_1
\tr(\sigma_1X)+\lambda_1\sigma_2\tr(\sigma_2X)+\lambda_3\sigma_3\tr(\sigma_3X)
\right],
\end{equation}
where $\sigma_\alpha$ denote the Pauli matrices. The real numbers $\lambda_1$ and $\lambda_3$ are the eigenvalues of $\Lambda$ to the eigenvectors given by the eigenvalue equations,
\begin{equation}
\Lambda[\sigma_1]=\lambda_1\sigma_1,\qquad 
\Lambda[\sigma_2]=\lambda_1\sigma_2,\qquad 
\Lambda[\sigma_3]=\lambda_3\sigma_3.
\end{equation}
Note that $\Lambda$ is in general non-unital ($\Lambda[\mathbb{I}]\neq\mathbb{I}$). Instead of the maximally mixed state $\mathbb{I}/2$, it preserves the state
\begin{equation}
\rho_\ast=\frac 12 \left[\mathbb{I}+\frac{\lambda_\ast}{1-\lambda_3}\sigma_3\right],
\end{equation}
meaning that $\Lambda[\rho_\ast]=\rho_\ast$. Therefore, the real number $\lambda_\ast$ determines the invariant state of $\Lambda$.

The positivity of the phase-covariant map is determined by its action on an arbitrary density matrix
\begin{equation}
\rho=\frac 12 \begin{pmatrix}
1+x_3 & x_1-ix_2 \\
x_1+ix_2 & 1-x_3
\end{pmatrix},\qquad x_1^2+x_2^2+x_3^2\leq 1.
\end{equation}
The output state is represented via
\begin{equation}
\Lambda[\rho]=\frac 12 \begin{pmatrix}
1+\lambda_\ast+\lambda_3x_3 & \lambda_1(x_1-ix_2) \\
\lambda_1(x_1+ix_2) & 1-\lambda_\ast -\lambda_3x_3
\end{pmatrix},
\end{equation}
whose eigenvalues
\begin{equation}
\mu_\pm=\frac 12 \left(1\pm\sqrt{\lambda_1^2(x_1^2+x_2^2)+(\lambda_\ast+\lambda_3x_3)^2}\right)
\end{equation}
are positive for any input state if and only if
\begin{equation}\label{P_cond}
|\lambda_1|\leq 1,\qquad |\lambda_3|+|\lambda_\ast|\leq 1,
\end{equation}
which give the positivity conditions for $\Lambda$. Now, $\Lambda$ is completely positive (and therefore, a quantum channel) provided that the associated Choi matrix
\begin{equation}\label{Choi}
\rho_\Lambda:=\frac 12 \sum_{k,\ell=0}^1|k\>\<\ell|\otimes\Lambda[|k\>\<\ell|]=
\frac{1}{4}\begin{pmatrix}
1+\lambda_\ast+\lambda_3 & 0 & 0 & 2\lambda_1 \\
0 & 1-\lambda_\ast-\lambda_3 & 0 & 0 \\
0 & 0 & 1+\lambda_\ast-\lambda_3 & 0 \\
2\lambda_1 & 0 & 0 & 1-\lambda_\ast+\lambda_3 
\end{pmatrix}
\end{equation}
is positive-semidefinite. From the form of its eigenvalues
\begin{equation}
\alpha_\pm=\frac 14 (1-\lambda_3\pm\lambda_\ast),\qquad
\beta_\pm=\frac 14 (1+\lambda_3\pm\sqrt{\lambda_\ast^2+4\lambda_1^2}),
\end{equation}
it is straightforward to show that complete positivity of $\Lambda$ is equivalent to \cite{Filippov}
\begin{equation}\label{CP_cond}
|\lambda_3|+|\lambda_\ast|\leq 1,\qquad 4\lambda_1^2+\lambda_\ast^2\leq(1+\lambda_3)^2.
\end{equation}
An important class of quantum maps are entanglement breaking channels, which produce separable states when acting on a composite system. It was shown that a qubit channel is entanglement breaking if and only if its Choi matrix satisfies \cite{qubitEBC}
\begin{equation}
\rho_\Lambda\leq \frac 12 \mathbb{I}\otimes\mathbb{I}.
\end{equation}
For phase-covariant channels, the above condition reduces to
\begin{equation}\label{EBC_cond}
4\lambda_1^2+\lambda_\ast^2\leq(1-\lambda_3)^2.
\end{equation}

Quantum channels are used not only to provide discrete evolution of physical systems, which occurs e.g. in quantum processing and quantum measurements \cite{TQI}. Continuous models describe time evolution of open quantum systems  by applying dynamical maps -- that is, families of time-parameterized quantum channels $\{\Lambda(t)|t\geq 0,\Lambda(0)=\oper\}$ \cite{BreuerPetr}. Dynamical maps are the solutions of master equations, the simplest being the dynamical equation for the Markovian semigroup
\begin{equation}\label{MS}
\dot{\Lambda}(t)=\mathcal{L}\Lambda(t),\qquad\Lambda(0)=\oper,
\end{equation}
with the time-independent Gorini-Kossakowski-Sudarshan-Lindblad (GKSL) generator $\mathcal{L}$ \cite{GKS,L}. For the phase-covariant dynamical maps \cite{Filippov},
\begin{equation}\label{L}
\mathcal{L}=\gamma_+\mathcal{L}_++\gamma_-\mathcal{L}_-+\gamma_3\mathcal{L}_3
\end{equation}
with the decoherence rates $\gamma_\pm(t)$, $\gamma_3(t)$ and
\begin{equation}
\mathcal{L}_{\pm}[X]=\sigma_\pm X\sigma_\mp -\frac 12 \{\sigma_\mp\sigma_\pm,X\},
\qquad \mathcal{L}_3[X]=\frac 14(\sigma_3X\sigma_3-X).
\end{equation}
Memory effects of quantum evolution are often included by replacing the constant generator $\mathcal{L}$ with the time-local generator $\mathcal{L}(t)$ of the same form but with time-dependent (not necessarily positive) rates $\gamma_\pm(t)$, $\gamma_3(t)$. The corresponding dynamical map is characterized by time-dependent eigenvalues \cite{Filippov}
\begin{equation}
\lambda_1(t)=\exp\left\{-\frac 12 \Big[\Gamma_+(t)+\Gamma_-(t)+\Gamma_3(t)\Big]\right\},\qquad
\lambda_3(t)=\exp\Big[-\Gamma_+(t)-\Gamma_-(t)\Big],
\end{equation}
\begin{equation}
\lambda_\ast(t)=\exp\Big[-\Gamma_+(t)-\Gamma_-(t)\Big]\int_0^t
\Big[\gamma_+(\tau)-\gamma_-(\tau)\Big]\exp\Big[\Gamma_+(\tau)+\Gamma_-(\tau)\Big]
\der\tau,
\end{equation}
where $\Gamma_k(t)=\int_0^t\gamma_k(\tau)\der\tau$, $k=\pm,3$. Note that a vanishing eigenvalue corresponds to a singular generator with at least one infinite decoherence rate, which is unphysical. Physical time-local generators always produce dynamical maps with
\begin{equation}\label{TLG_cond}
\lambda_1(t)>0,\qquad\lambda_3(t)>0.
\end{equation}
Finally, quantum maps $\Lambda$ arise from dynamical maps $\Lambda(t)$ with a fixed time $t=t_\ast\geq 0$. We refer to such channels as {\it obtainable with time-local generators}.

\section{Line and volume elements}

In this section, we follow the method from ref. \cite{Pauli_volume} to introduce geometrical structures in the space of Choi states associated with the phase-covariant maps. Recall that, due to the Choi-Jamio{\l}kowski isomorphism \cite{Choi, Jamiolkowski}, there exists a one-to-one correspondence between quantum maps $\Lambda$ and quantum states $\rho_\Lambda$ (see eq. (\ref{Choi})). From now on, we work on the space of the Choi states. We equip it with the Hilbert-Schmidt metric $g=\frac 14 \mathrm{diag}(2,1,1)$ defined by the line element
\begin{equation}\label{HS}
\der s^2:=\Tr(\der\rho_\Lambda^2)=\frac 14 (2\der\lambda_1^2+\der\lambda_3^2+\der\lambda_\ast^2).
\end{equation}
This allows us to introduce the corresponding volume element
\begin{equation}\label{dV}
\der V:=\sqrt{\det g}\der\lambda_1\der\lambda_2\der\lambda_3=
\frac{\sqrt{2}}{8}\der\lambda_1\der\lambda_3\der\lambda_\ast.
\end{equation}
Interestingly, on the level of geometrical structures, the parameter $\lambda_\ast$ that is responsible for non-unitality of $\Lambda$ behaves just like the channel eigenvalues $\lambda_1$ and $\lambda_3$.

Comparing our results to the geometry of Pauli channels
\begin{equation}
\Lambda[\sigma_\alpha]=\lambda_\alpha\sigma_\alpha,\qquad\alpha=0,1,2,3,\qquad,
\lambda_0=1,
\end{equation}
for which
\begin{equation}\label{HS2}
\der s^2=\frac 14 (\der\lambda_1^2+\der\lambda_2^2+\der\lambda_3^2),
\qquad
\der V=\frac 18 \der\lambda_1\der\lambda_2\der\lambda_3,
\end{equation}
one observes certain similarities. Both metrics are diagonal in $\lambda_\alpha$'s. The difference in the first term of $g$ comes from the fact that $\lambda_1$ for the phase-covariant channels is two-times degenerate, while all the other parameters are non-degenerate. This also influences the final formula for the volume element. By fixing $\lambda_\ast=0$ in eq. (\ref{HS}) and $\lambda_2=\lambda_1$ in eq. (\ref{HS2}), we recover the common subspace for the Pauli channels and phase-covariant channels.

Now, the volume element from eq. (\ref{dV}) can be used to determine volumes of various classes of the phase-covariant maps. These volumes
\begin{equation}
V=\int_{\mathcal{C}_{\mathrm{M}}}\der V
\end{equation}
are calculated by integrating $\der V$ over the corresponding regions. In particular, we are interested in the following regions of integration:
\begin{enumerate}[(i)]
\item the positivity region given by eq. (\ref{P_cond}),
\begin{equation}
\mathcal{C}_{\mathrm{PT}}=\left\{\lambda_1,\lambda_3,\lambda_\ast:\,
|\lambda_1|\leq 1\,\wedge\,|\lambda_3|\leq 1\,\wedge\,|\lambda_3|+|\lambda_\ast|\leq 1\right\},
\end{equation}
\item the complete positivity region from eq. (\ref{CP_cond}),
\begin{equation}
\mathcal{C}_{\mathrm{CPT}}=\left\{\lambda_1,\lambda_3,\lambda_\ast:\,
|\lambda_3|\leq 1\,\wedge\,|\lambda_3|+|\lambda_\ast|\leq 1\,\wedge\, 4\lambda_1^2+\lambda_\ast^2\leq(1+\lambda_3)^2\right\},
\end{equation}
\item the entanglement breaking region in eq. (\ref{EBC_cond}),
\begin{equation}
\mathcal{C}_{\mathrm{EBC}}=\left\{\lambda_1,\lambda_3,\lambda_\ast:\,
|\lambda_3|\leq 1\,\wedge\,|\lambda_3|+|\lambda_\ast|\leq 1\,\wedge\, 
4\lambda_1^2+\lambda_\ast^2\leq(1\pm\lambda_3)^2\right\},
\end{equation}
\item the region of positive eigenvalues that corresponds to maps obtainable with physical time-local generators (see eq. (\ref{TLG_cond})),
\begin{equation}
\mathcal{C}_{\mathrm{TLG}}=\left\{\lambda_1,\lambda_3,\lambda_\ast:\,
\lambda_1>0\,\wedge\,\lambda_3>0\,\wedge\,|\lambda_\ast|\leq 1\right\}.
\end{equation}
\end{enumerate}

\section{Volume of non-unital maps}

\begin{figure}[ht!]
  \includegraphics[width=0.5\textwidth]{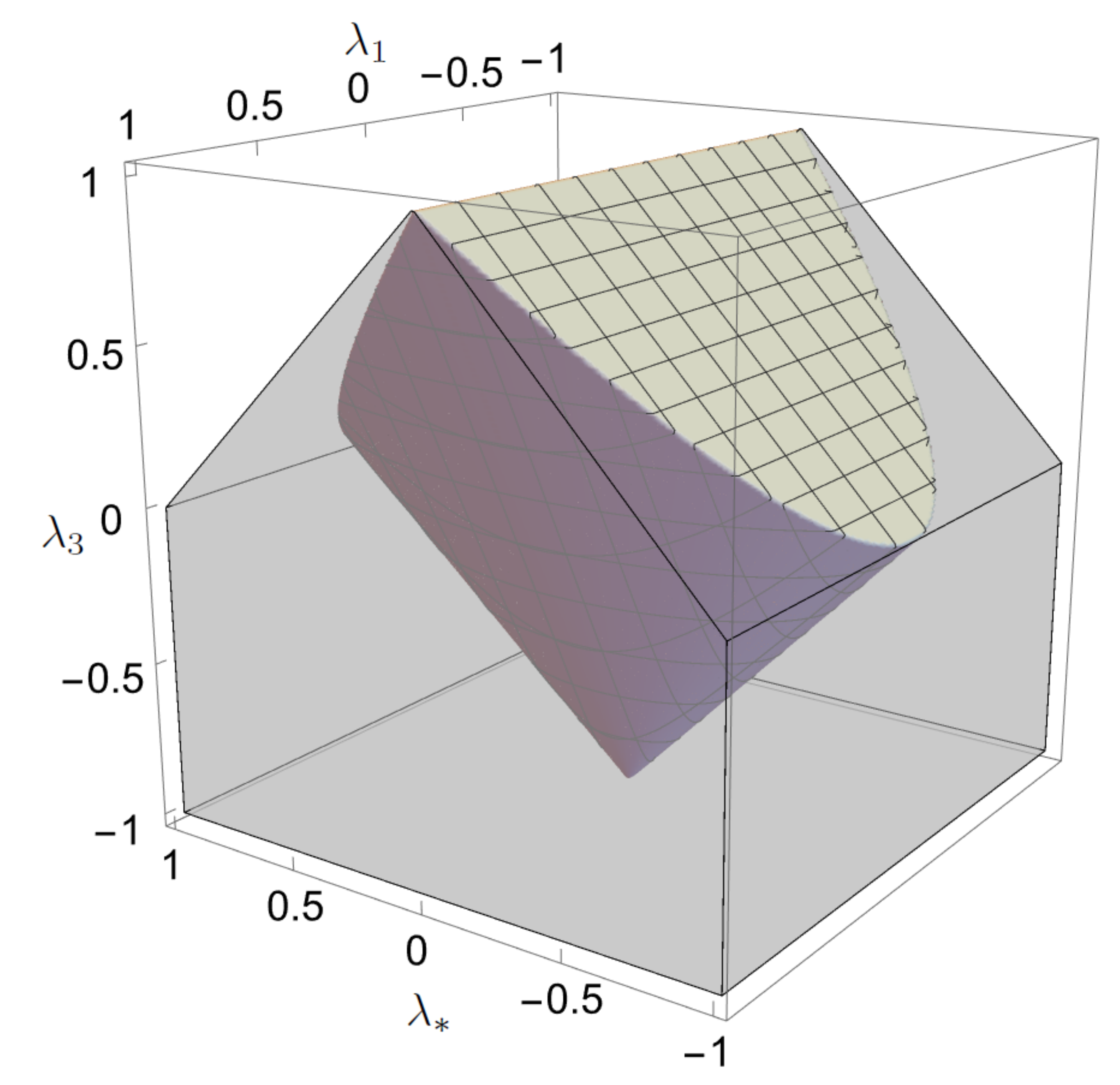}
\caption{
The range of parameters $\lambda_1,\ \lambda_3,\ \lambda_\ast$ corresponding to the positive, trace-preserving phase-covariant maps (light gray) and the phase-covariant channels (meshed on the surface of the PT region, dark gray inside of it).}
\label{PCP}
\end{figure}

Integrating the volume element $\der V$ over the positivity $\mathcal{C}_{\mathrm{P}}$ and complete positivity $\mathcal{C}_{\mathrm{CP}}$ regions, one arrives at the volumes
\begin{equation}\label{ebc}
V(\mathcal{C}_{\rm PT})=\frac{\sqrt{2}}{2},\qquad V(\mathcal{C}_{\rm CPT})=\frac{2\sqrt{2}}{9}
\end{equation}
of the positive, trace-preserving phase-covariant maps (PT) and channels (CPT), respectively. Note that $V(\mathcal{C}_{\rm CPT})/V(\mathcal{C}_{\rm PT})=4/9$, so less than a half of PT maps are CPT. This ratio is close to the relative volume for the Pauli maps with $\lambda_2=\lambda_1$ (or phase-covariant channels with $\lambda_\ast=0$), where the channels amount to exactly a half of the positive, trace-preserving maps \cite{PV}. The shapes of $\mathcal{C}_{\rm PT}$ and $\mathcal{C}_{\rm CPT}$ regions are plotted in Fig. \ref{PCP}. It is easy to see that they are more complicated than the PT and CPT regions for the Pauli channels (a cube and a tetrahedron, respectively). In particular, $\mathcal{C}_{\rm CPT}$ resembles a house with a square floor, while $\mathcal{C}_{\rm PT}$ is an elliptic cone with a vertix at $(\lambda_1,\lambda_3,\lambda_\ast)=(0,-1,0)$, cut off by the house's roof.

\begin{figure*}
\centering
\begin{minipage}[b]{.4\textwidth}
  \includegraphics[width=\textwidth]{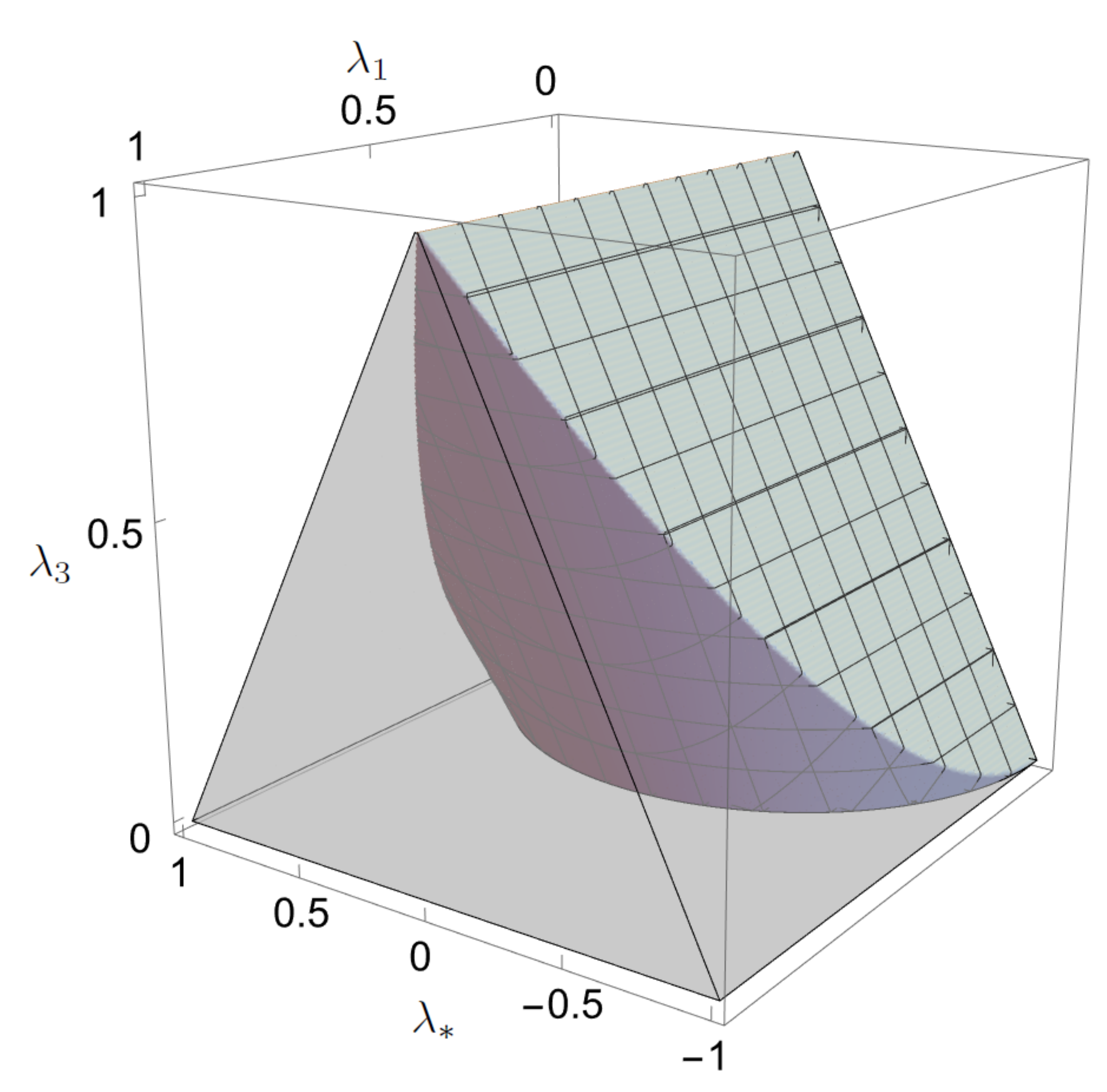}
\caption{
A graphical representation of the range of the eigenvalues $\lambda_1,\ \lambda_2,\ \lambda_3$ that describe the Pauli channels (light gray) and their subclass that is obtainable with time-local generators (dark gray).}
\label{TLG_CPT}
\end{minipage}\qquad
\begin{minipage}[b]{.4\textwidth}
  \includegraphics[width=\textwidth]{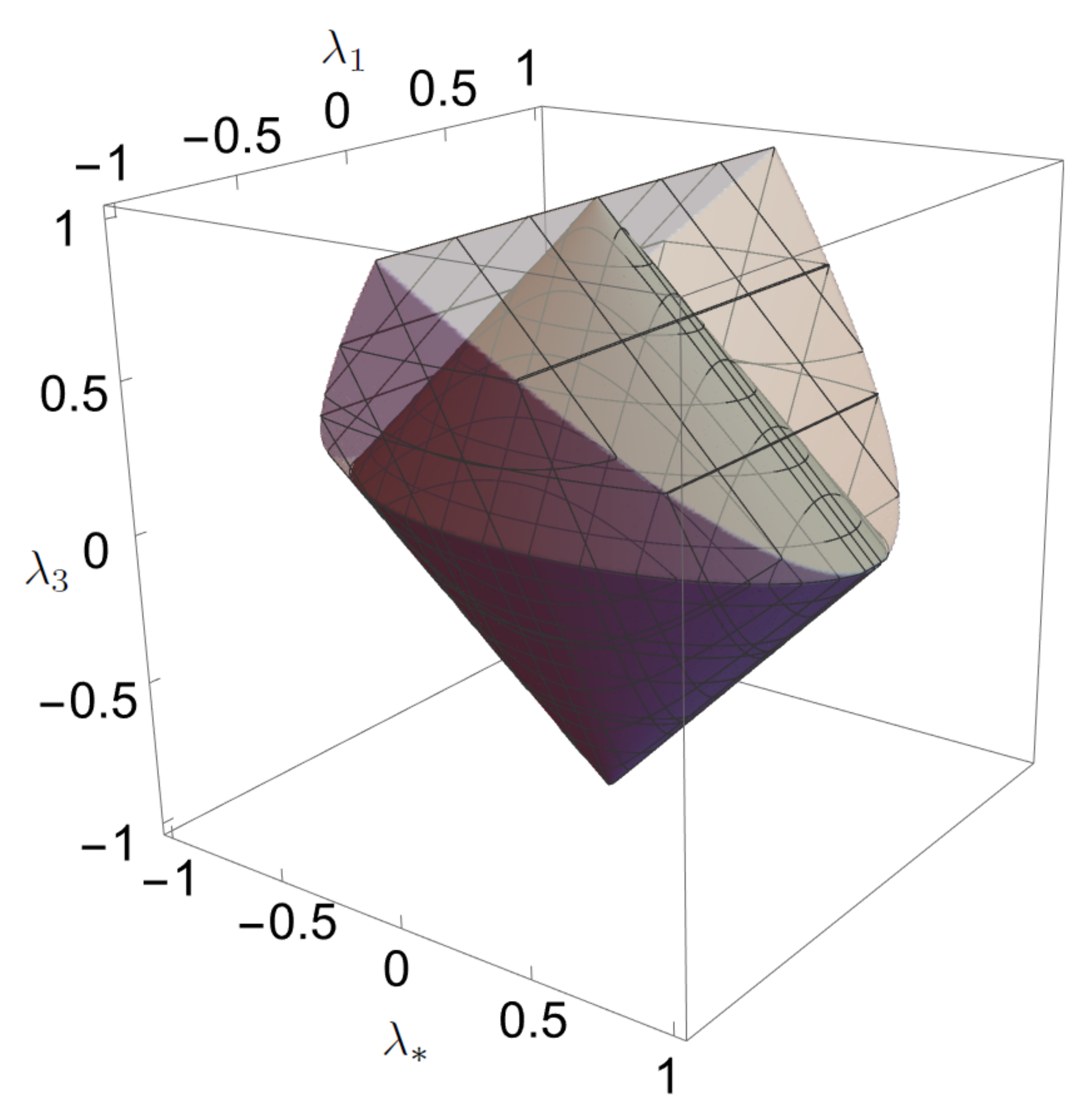}
\caption{
The channel parameters range for the entanglement breaking channels (dark gray) inside the complete positivity region (light gray). For $\lambda_3\leq 0$, both regions coincide.}
\label{EBC}
\end{minipage}
\end{figure*}

Among the phase-covariant maps, there are some that can be obtained with time-local generators (TLG). For positive, trace-preserving maps, the relative volume reads
\begin{equation}
\frac{V(\mathcal{C}_{\rm PT}\cap\mathcal{C}_{\rm TLG})}{V(\mathcal{C}_{\rm PT})}=\frac 14,
\end{equation}
whereas for the quantum channels
\begin{equation}
\frac{V(\mathcal{C}_{\rm CPT}\cap\mathcal{C}_{\rm TLG})}{V(\mathcal{C}_{\rm CPT})}=\frac 12 - \frac{3\pi}{64}\simeq 0.35.
\end{equation}
Wwe graphically represent the ranges of $\lambda_1$, $\lambda_3$, $\lambda_\ast$ that correspond to these regions in Fig. \ref{TLG_CPT}. The region $\mathcal{C}_{\rm PT}\cap\mathcal{C}_{\rm TLG}$ is a triangular prism (light gray), whereas $\mathcal{C}_{\rm CPT}\cap\mathcal{C}_{\rm TLG}$ is a cutoff of the elliptic cone (dark gray inside the prism, meshed on its faces). The shape in the $\lambda_1\lambda_\ast$-plane is a half-ellipse, and in the $\lambda_3\lambda_\ast$-plane, there is an isosceles triangle. The remaining fraction of maps -- that is,
\begin{equation}
1-\frac{V(\mathcal{C}_{\rm CPT}\cap\mathcal{C}_{\rm TLG})}{V(\mathcal{C}_{\rm CPT})}=\frac 12 + \frac{3\pi}{64}\simeq 0.65,
\end{equation}
corresponds to the volume ratio of the phase-covariant channels that are obtainable only by considering the non-local master equations \cite{Nakajima,Zwanzig}
\begin{equation}\label{K}
\dot{\Lambda}(t)=\int_0^tK(t,\tau)\Lambda(\tau)\der\tau,\qquad\Lambda(0)=\oper,
\end{equation}
with memory kernels $K(t,\tau)$. As only around a third of all channels arise from time-local master equations with regular generators, these results show us how important it is to further develop the theories of non-invertible dynamical maps, admissible memory kernels, and singular generators.

%

Among all the phase-covariant channels, there are
\begin{equation}
\frac{V(\mathcal{C}_{\rm EBC})}{V(\mathcal{C}_{\rm CPT})}=\frac{3\pi}{16}\simeq 0.59
\end{equation}
channels that break quantum entanglement, which interestingly is more than $1/2$ for the Pauli maps with $\lambda_2=\lambda_1$ \cite{PV}.
Finally, we can ask how much of the phase-covariant channels with positive eigenvalues break quantum entanglement. The corresponding volume ratio is
\begin{equation}
\frac{V(\mathcal{C}_{\rm EBC}\cap\mathcal{C}_{\rm TLG})}{V(\mathcal{C}_{\rm CPT}\cap\mathcal{C}_{\rm TLG})}=\frac{3\pi}{32-3\pi}\simeq 0.42.
\end{equation}
Moreover, among all the entanglement breaking channels,
\begin{equation}
\frac{V(\mathcal{C}_{\rm EBC}\cap\mathcal{C}_{\rm TLG})}{V(\mathcal{C}_{\rm EBC})}=\frac 14
\end{equation}
can be obtained with time-local generators. Our results are presented in Fig. \ref{EBC}.It shows that the entanglement breaking region (dark gray) is inscribed inside the complete positivity region (light gray). Observe that $\mathcal{C}_{\rm EBC}$ is in the shape of an elliptic bicone, which bears certain symmetry resemblence to the EBC region of the Pauli channels (octahedron). The lower half of the bicone is coplanar with the CPT region, whereas its upper half has common boundary with $\mathcal{C}_{\rm CPT}$ at $\lambda_1=0$.

\begin{figure}[ht!]
  \includegraphics[width=0.75\textwidth]{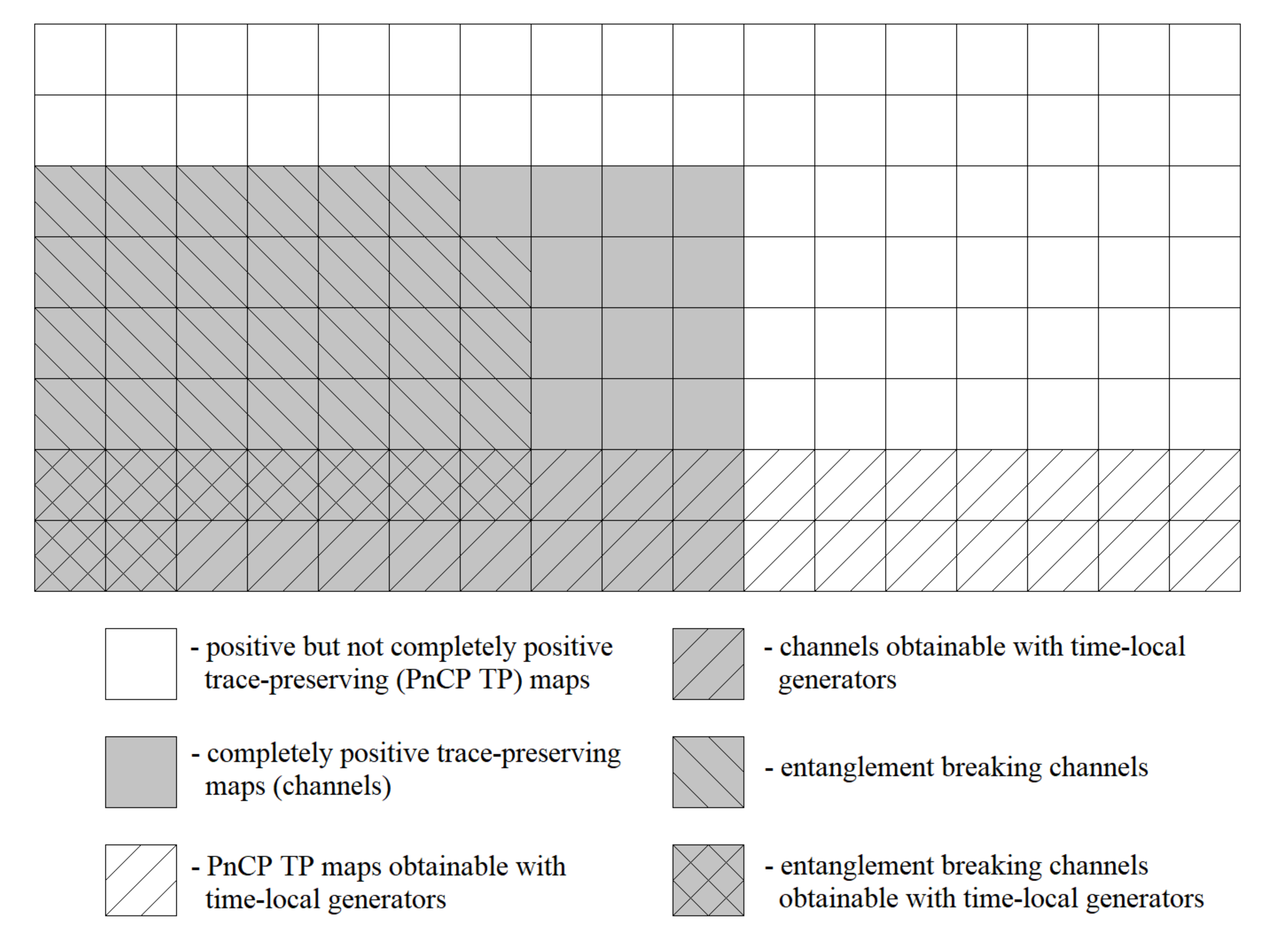}
\caption{An approximate quantitative representation of the volumes for special classes of trace-preserving phase-covariant maps.}
\label{pie}
\end{figure}

The results of this section are summarized in Fig. \ref{pie}. The rectangle divided into $8\times 17$ identical squares represents the volume of positive, trace-preserving phase-covariant maps $V(\mathcal{C}_{\mathrm{PT}})$. The gray area shows the part occupied by the quantum channels, whereas the white area corresponds to the positive but not completely positive maps. The hatched regions are associated with the volumes of entanglement breaking channels (left-to-right) and the positive maps obtainable with time-local generators (right-to-left). The smallest part of the rectangle corresponds to $\mathcal{C}_{\rm EBC}\cap\mathcal{C}_{\rm TLG}$, which is not surprising, as these maps satisfy the most restrictive conditions. Interestingly, the next smallest regions are $(\mathcal{C}_{\rm CPT}\cap\mathcal{C}_{\rm TLG})\setminus\mathcal{C}_{\rm EBC}$ and $(\mathcal{C}_{\rm CPT}\setminus\mathcal{C}_{\rm TLG})\setminus\mathcal{C}_{\rm EBC}$, respectively. On the other hand, the biggest area is occupied by positive, trace-preserving maps obtainable only with memory kernels.

Even though the formula for the line element is general, the volume element is not. For example, the volume for a fixed $\lambda_\alpha$ is zero -- unless we restrict the manifold to a hypersurface. This way, we can calculate relative volumes for unital maps ($\lambda_\ast=0$), quantum maps with even more degenerated eigenvalues ($\lambda_1=\lambda_3$), or even non-invertible mappings ($\lambda_1=0$ or $\lambda_3=0$). For a full analysis of unital qubit maps ($\lambda_\ast=0$), refer to our previous works \cite{Pauli_volume,PV}.

\subsection{Symmetric maps}

By introducing additional symmetries, we consider the isotropic maps with a three-times degenerated eigenvalue $\lambda_1=\lambda_3$. In this case, the phase-covariant maps are fully determined by only two real parameters: $\lambda_3$ and $\lambda_\ast$. Now, the line and volume elements are given by
\begin{equation}
\der s^2=\frac 14 (3\der\lambda_3^2+\der\lambda_\ast^2),\qquad\der V=\frac {\sqrt{3}}{4} \der\lambda_3\der\lambda_\ast.
\end{equation}
The regions corresponding to special classes of phase-covariant maps are shown in Fig. \ref{symm}.a. Observe that there are as many channels obtainable with time-local generators that are not entanglement breaking as positive but not completely positive TP maps. Moreover, exactly half of entanglement breaking channels is only obtainable with memory kernels. Relatively to positive, trace-preserving maps, there are
\begin{equation}
\frac{V(\mathcal{C}_{\mathrm{CPT}})}{V(\mathcal{C}_{\mathrm{PT}})}
=\frac{9+2\pi\sqrt{3}}{27}\simeq 0.74
\end{equation}
quantum channels, whereas entanglement breaking channels constitute
\begin{equation}
\frac{V(\mathcal{C}_{\mathrm{EBC}})}{V(\mathcal{C}_{\mathrm{CPT}})}
=\frac{4\pi\sqrt{3}-9}{2\pi\sqrt{3}+9}\simeq 0.20
\end{equation}
of all phase-covariant channels.


If we include one last symmetry constraint, which is taking $\lambda_\ast=\lambda$, then $\der s^2=\der\lambda^2$ and $\der V=\der\lambda$.
In this case, the regions of integration reduce to
\begin{align*}
\begin{split}
\mathcal{C}_{\mathrm{PT}}&=\left\{\lambda:\,|\lambda|\leq\frac 12\right\},\\
\mathcal{C}_{\mathrm{CPT}}&=\left\{\lambda:\,-\frac 14 (\sqrt{5}-1)\leq\lambda\leq\frac 12\right\},
\end{split}\qquad
\begin{split}
\mathcal{C}_{\mathrm{EBC}}&=\left\{\lambda:\,|\lambda|\leq \frac 14 (\sqrt{5}-1)\right\},\\
\mathcal{C}_{\mathrm{PT}}\cap\mathcal{C}_{\mathrm{TLG}}&=
\mathcal{C}_{\mathrm{CPT}}\cap\mathcal{C}_{\mathrm{TLG}}=\left\{\lambda:\,0\leq\lambda\leq \frac 12\right\}.
\end{split}
\end{align*}
Now, nearly three out of four positive TP maps are completely positive. Exactly a half of PTP maps and over two thirds of channels are obtainable with time-local generators. Almost one in every five channels breaks quantum entanglement. Similarly to the less symmetric case, there are as many entanglement breaking channels obtainable with time-local generators as with only memory kernels.


\subsection{Non-invertible maps}

\begin{figure}[htb!]
  \includegraphics[width=\textwidth]{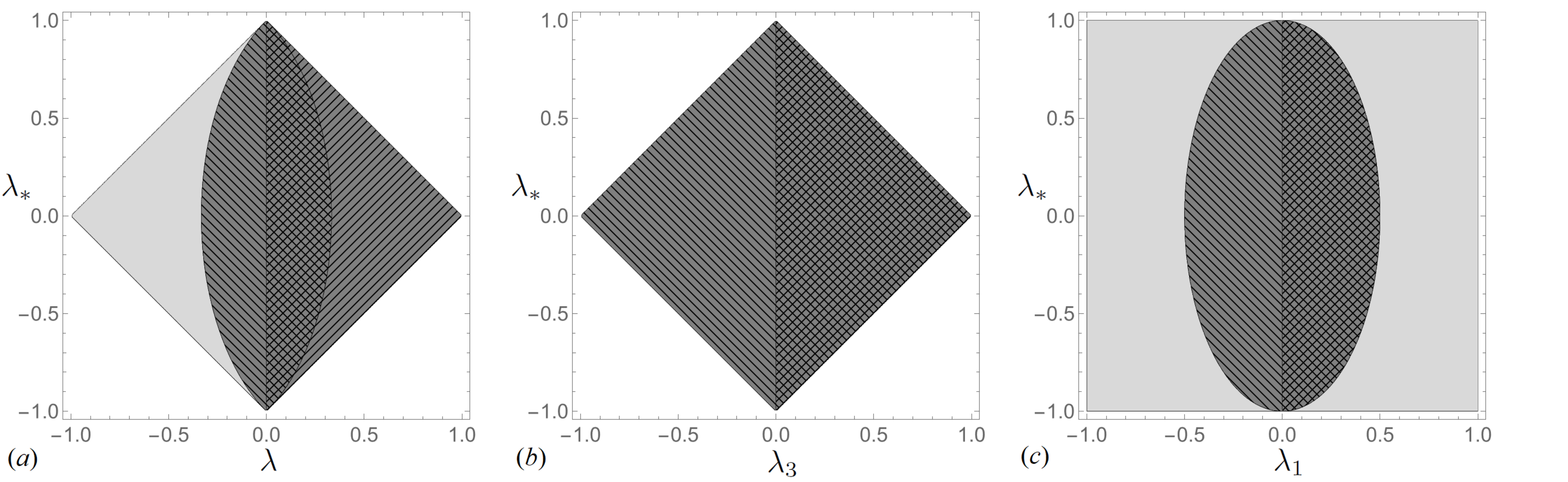}
\caption{
Admissible ranges of the map parameters that correspond to the isotropic maps with $\lambda_1=\lambda_3\equiv\lambda$ (a), as well as non-invertible maps with $\lambda_1=0$ (b) or $\lambda_3=0$ (c). The light gray areas denote positivity regions, the dark gray areas are associated with complete positivity regions, and the hatching indicates the regions of entanglement breaking (left-to-right) and positive eigenvalues (right-to-left).}
\label{symm}
\end{figure}

For a map to be non-invertible, at least one of its eigenvalues has to vanish. However, if $\lambda_1=\lambda_3=0$, then every positive TP map is trivially an entanglement breaking channel not obtainable with time-local generators. Therefore, we first focus on $\lambda_1=0$ and $\lambda_3\neq 0$. Now, the geometry of the hypersurface is determined by
\begin{equation}
\der s^2=\frac 14 (\der\lambda_3^2+\der\lambda_\ast^2),\qquad\der V=\frac 14 \der\lambda_3\der\lambda_\ast.
\end{equation}
By examining the regions of integration, we see that
\begin{align*}
\mathcal{C}_{\mathrm{PT}}&=\mathcal{C}_{\mathrm{CPT}}=\mathcal{C}_{\mathrm{EBC}}
=\left\{\lambda_3,\lambda_\ast:\,|\lambda_3|\leq 1,\,|\lambda_3|+|\lambda_\ast|\leq 1\right\},\\
\mathcal{C}_{\mathrm{PT}}\cap\mathcal{C}_{\mathrm{TLG}}&=
\left\{\lambda_3,\lambda_\ast:\,0\leq\lambda_3\leq 1,\,|\lambda_\ast|\leq 1-\lambda_3\right\}.
\end{align*}
Our observations agree with Fig. \ref{symm}.b, where there are only two visible, equinumerous regions of PTP maps: entanglement breaking with positive and negative eigenvalue $\lambda_3$.

Let us observe what happens if one instead considers the maps with $\lambda_3=0$ and $\lambda_1\neq 0$. Unlike in the case of Pauli channels, the positivity, complete positivity, and entanglement breaking conditions are not symmetric with respect to $\lambda_1\leftrightarrow\lambda_3$. This is reflected in the geometry of the manifold, as now we have
\begin{equation}
\der s^2=\frac 14 (2\der\lambda_1^2+\der\lambda_\ast^2),\qquad\der V=\frac{\sqrt{2}}{4} \der\lambda_1\der\lambda_\ast,
\end{equation}
and also
\begin{align*}
\mathcal{C}_{\mathrm{PT}}&=\left\{\lambda_1,\lambda_\ast:\,|\lambda_1|\leq 1,\,|\lambda_\ast|\leq 1\right\},\\
\mathcal{C}_{\mathrm{CPT}}&=\mathcal{C}_{\mathrm{EBC}}=
\left\{\lambda_1,\lambda_\ast:\,|\lambda_\ast|\leq 1,\,|\lambda_1|\leq\frac{1-\lambda_\ast^2}{2}\right\},\\
\mathcal{C}_{\mathrm{TLG}}&=\left\{\lambda_1,\lambda_\ast:\,0\leq\lambda_1\leq 1,\,|\lambda_\ast|\leq 1\right\}.
\end{align*}
The integration regions are presented in Fig. \ref{symm}.c. Finally, the volume of positive, trace-preserving maps $V(\mathcal{C}_{\rm PT})=4$ is bigger than that of quantum channels $V(\mathcal{C}_{\rm CPT})=\pi/2$, similarly to the case of three-parameter phase-covariant maps in eq. (\ref{ebc}). Just like in Fig. \ref{symm}.b, we again have the complete positivity region that has two axes of symmetry. Once more, every single channel is entanglement breaking.

%

\section{Conclusions}

In this paper, we consider the geometrical aspects of the space of phase-covariant maps. We analytically derive the regions corresponding to positive, completely positive, entanglement breaking, and time-local generated maps. For these special classes of maps and channels, we use the Hilbert-Schmidt metric on the manifold of Choi-Jamio{\l}kowski states to analytically calculate their volumes. We compare our results with the volumes of Pauli channels. The calculations are then repeated for the hypersurfaces of isotropic and non-invertible maps.

Relative volumes of quantum maps can be interpreted as a probability of randomly selecting a phase-covariant channel with the desired properties. Interestingly, even if we know nothing about the channel eigenvalues, we still possess some partial knowledge about the relative volumes. The more symmetries are introduced, the more channels satisfy the additional constraints of being entanglement breaking or possessing strictly positive eigenvalues. Despite that, there is still a huge amount of channels not realized via time-local master equations, which indicates the importance of further developing the memory kernel approach to quantum evolution.

The geometry of non-unital channels is still a relatively unexplored area of research. Therefore, there are many open questions that require further study. First of all, it would be interesting to consider more general quantum maps, both for qubit and qudit evolution. One could try to determine the amounts of channels using different measures, like the Fisher-Rao measure, the Lebesque measure, or the Haar measure. Finally, one could ask how the relative volumes of dynamical maps evolve in time under different types of quantum evolution.

\section{Acknowledgements}

This research was funded in whole or in part by the National Science Centre, Poland, Grant number 2021/43/D/ST2/00102. For the purpose of Open Access, the author has applied a CC-BY public copyright licence to any Author Accepted Manuscript (AAM) version arising from this submission.

\bibliography{C:/Users/cyndaquilka/OneDrive/Fizyka/bibliography}
\bibliographystyle{C:/Users/cyndaquilka/OneDrive/Fizyka/beztytulow2}

\end{document}